# Altermagnetism Classification

**Sang-Wook Cheong[*,1] and Fei-Ting Huang[1]**

[1]Keck Center for Quantum Magnetism and Department of Physics and Astronomy, Rutgers University, Piscataway, NJ 08854, USA

*Corresponding author: sangc@physics.rutgers.edu

**Altermagnets are defined as magnetic states with fully compensated spin angular momenta (spins) and broken PT (P: parity, T: time reversal, PT: parity times time reversal) symmetry. We classify three kinds of altermagnets: M-type: broken T symmetry and nonzero net magnetic moments, S-type: broken T symmetry and zero net magnetic moments, and A-type: unbroken T and broken P symmetries. Furthermore, strong altermagnets have spin split bands through exchange coupling in the non-relativistic limit, *i.e.,* for zero spin-orbit coupling (SOC), and weak altermagnets has spin split bands only with non-zero SOC. These strong vs. weak altermagnets can be identified from the total number of symmetric orthogonal spin rotation operations. These classifications of altermagnets will be an essential guidance for the future research on altermagnetism.**

## INTRODUCTION

Altermagnetism[1-7] has been introduced as (almost) antiferromagnetism having time-reversal symmetry breaking by combining 'centrosymmetric crystallographic structure with local binary structural alternation' and 'collinear antiferromagnetism with time reversal and spatial inversion symmetries', even when translation is freely allowed. These altermagnets have broken **PT** (**P**: parity, **T**: time reversal, **PT**: parity times time reversal)[8] symmetry and split bands[1,2], even in the non-relativistic limit, i.e. for zero spin-orbit coupling (SOC). The concept of altermagnetism has been extended to non-collinear (almost) antiferromagnets with multiple local structural variants and spin orientations[8-11]. These altermagnets with spin-split bands exhibit numerous spin-relevant phenomena such as various-order anomalous Hall effects[12-14], piezomagnetism[8,9,15,16], kinetomagnetism (i.e., current-induced magnetization)[17], etc. In this work, we broaden the definition of altermagnetism by focusing on the existence of spin-split bands rather than just structural alterations. We also include altermagnets that break spatial inversion symmetry while preserving time-reversal symmetry. The influence of spin-orbital coupling is categorized into weak (relativistic) and strong (non-relativistic) altermagnets. All strong collinear altermagnets discussed in the original proposal[3,5] in the literature are a subset of our M-type and S-type strong altermagnets. Our refined definition of altermagnetism is: magnetism with broken **PT** symmetry and complete spin compensation in the non-relativistic limit. This refined definition encompasses all instances of the original concept of altermagnetism, including those with collinear spins and alternating structural variations. Furthermore, we clarify the concept of 'almost' antiferromagnetism, introduce a classification scheme for three types of altermagnets (M-, S-, and A-types), and differentiate between strong and weak altermagnets through the consideration of straightforward spin rotation symmetry, utilizing the fundamental concept of the spin space/point group[18-20].

## SAM ALTERMAGNETISM CLASSIFICATION

Figure 1a illustrates **PT** symmetry[21] operation on '+$k$ with spin up' links '+$k$ with spin down'. This means that the requirement for a system to have a spin-split bands is broken **PT** symmetry. We first define 'altermagnets as magnetic states with fully compensated spin angular momenta (spins) and broken **PT** symmetry'. We focus on physical properties that are invariant to any translation, so we will always allow free translations. In other words, we always consider symmetry operations with a combination of any translation. For example, we will argue that the simple collinear antiferromagnetic state in Fig. 1c has unbroken **T** symmetry, while the ferromagnetic state is Fig. 1b still has broken **T** symmetry. There are two ways to have broken **PT** symmetry: [A] broken **T** symmetry and [B] unbroken **T** symmetry but broken **P** symmetry. Since we are discussing broken **PT** symmetry, we exclude the case of 'broken **T** symmetry, broken **P** symmetry and unbroken **PT** symmetry'. Also note that both 'broken **T**, unbroken **P** and broken **PT**', and 'broken **T**, broken **P** and broken **PT**' are parts of [A]. It turns out that Magnetization (*M*) along $x$ has broken {**PT**,**T**,$m_y$,$m_z$,$C_{2y}$,$C_{2z}$,$C_{3y}$,$C_{3z}$} with free rotation along $x$.[22] All magnetic point groups (MPGs), belonging the ferromagnetic point group, do have same or lower symmetry than *M*, and are a subset of [A]. In our discussion, spin, spin angular momentum, and spin (magnetic) moment are interchangeable, similarly orbital angular momentum and orbital (magnetic) moment are all interchangeable, and magnetic moment means the total magnetic moment combining spin and orbital moments.

Thus, we have well-defined three types of altermagnets with fully compensated spin angular momenta (spins) and broken **PT** symmetry: [1] M-type: Broken **T** symmetry and belonging to the ferromagnetic point group, exhibiting SOC-induced spin splitting at the Γ point. There exists orbital ferrimagnetism, and uncompensated 'M'agnetization in the ground state. M-type is classified as type-I in our previous report[8]. [2] S-type: Broken **T**, no net magnetization, and "S"ymmetric spin splitting, resulting in transverse piezomagnetism. S-type is classified as part of type-II in our previous report[8]. [3] A-type: Broken **P**, unbroken **T**, no net magnetization, and "A"ntisymmetric spin splitting, resulting in no transverse piezomagnetism, but potentially having quadratic (even-order) AHE. Many A-type altermagnets are candidates for E-switchable altermagntism, featuring unbroken **T** symmetry, but broken **P** symmetry. S/A-type: Broken **P**, broken **T**, no net magnetization, and both "S"ymmetric and "A"ntisymmetric spin splitting. S/A-type altermagnets exhibit the properties of S-type altermagnets as well as those of A-type altermagnets. The advantage of SAM altermagnetism classification lies in its connection to various kinetomagnetism[17], which are based on broken **PT** symmetry. For instance, electric current can induce magnetization in various order (linear, high-odd order, or even order)[23], and this induced magnetization can be associated with various anomalous Hall effect (AHE) or piezomagnetism[9,12,15,16,24] as discussed in the following sections.

**M-type and S-type altermagnets**

M-type altermagnets exhibit full spin compensation, leading to their ferromagnetic behavior primarily governed by orbital angular momenta, which originates from SOC. Thus, the meaning of 'almost' antiferromagnetism mentioned in the introduction is associated with these possible net magnetizations or magnetic moments due to orbital angular momenta in M-type altermagnets when SOC is considered. An intriguing example of an M-type altermagnet, exhibiting orbital ferrimagnetism, is $CoMnO_3$[25,26], where $Co^{2+}$ and $Mn^{4+}$ have the same spins, but the opposite sign of SOC. When non-zero orbital angular momenta develop through SOC, then the full spin compensation can become partially broken, and spins can also result in non-zero net values through SOC, which is a supplementary effect. Emphasize that the symmetry of M-type altermagnets does not warrantee zero net spin moment. Rather than that, even if their spin configurations with broken **PT** symmetry have zero net spin moment (e.g., collinear antiferromagnetic spins), their net magnetization can be non-zero.

Three examples of M-type altermagnets with broken **T** and **PT** symmetry are shown in Figs. 1d-1f. Fig. 1e corresponds to the magnetic structure of $NiF_2$[27,28], while Fig. 1f illustrates the basic unit of $Mn_3Sn$.[27,29] All spins in Figs. 1d-f are fully compensated, at least, for zero SOC. The magnetic state in Fig. 1d can have a non-zero net orbital moment along *x* due to *g*-tensor anisotropy, and that in Fig. 1e can have a canted magnetic moment along *y* through the Dzyaloshinskii-Moriya (DM) interaction – this canted moment is mostly an orbital moment, but spins can also contribute through SOC. Figure 1g displays an example of S-type altermagnets with broken **T** and **PT** symmetry, representing the magnetic structure of $CoF_2$ or $MnF_2$.[30,31] Magnetic structures of $NiF_2$ and $CoF_2$ are illustrated in the Supplementary Note 1.

M-type altermagnets (Fig. 1d-1f) with broken **T** and **PT** and non-zero magnetization, can exhibit linear AHE. S-type altermagnets with broken **T** and **PT** and zero magnetization can have transverse even-order current-induced magnetization (i.e., high-odd-order AHE). Thus, both M-type and S-type altermagnets can have transverse piezomagnetism. For example, the S-type altermagnet with **4'/mm'm** in Fig. 1g can exhibit transverse even-order current-induced magnetization along *z* with current along *x* or *y* (i.e., high-odd-order AHE with current along *x* (*y*) and Hall voltage along *y* (*x*)). Thus, it also shows transverse piezomagnetism with strain along *x* or *y* and induced magnetization along *z*.

**A-type altermagnets**

The cycloidal spin state in Fig. 1h accompanies electric polarization along *y* and the up-up-down-down spin state with two types of spins in Fig. 1j is associated with electric polarization along *x* [32]. Helical spins in Fig. 1i have chirality, i.e., broken all mirror symmetries. All of Figs. 1h–1j with fully compensated spins have unbroken **T** and broken **P**, so belong to A-type altermagnets. A-type altermagnets with broken **P** and **PT** can exhibit transverse odd-order current-induced magnetization (i.e., even-order AHE), and some M-type and all S/A-type altermagnets do have broken all of **P**, **T** and **PT**, so they can also show transverse even-order as well as odd-order current-induced magnetization (i.e., odd-order as well as even-order AHE). For instance, the A-type altermagnet with a spiral spin structure in MPG **m2m1'** in Fig. 1h can show transverse odd-order current-induced magnetization along *z* (*x*) with current along *x* (*z*) (i.e., even-order AHE with current along *x* or *z* and Hall voltage along *y*).

The chiral A-type altermagnet with **41'** in Fig. 1i can show longitudinal current-induced magnetization along any direction. Cycloidal-spin-order-driven multiferroicity seen in materials like orthorhombic $TbMnO_3$ [33] and $LiCu_2O_2$[34], as well as pyroelectric $Ni_2Mo_3O_8$ [35,36] show similar spin structures. The A-type altermagnet with **2mm1'** in Fig. 1j can show transverse odd-order current-induced magnetization along *z* (*y*) with current along *y* (*z*) (i.e., even-order AHE with current along *y* or *z* and Hall voltage along *x*). Materials such as $MnS_2$ [37], $HoMnO_3$ [38] show similar spin structures. Incommensurate (cycloidal/helical) spin modulations, including odd-period spin modulations, usually accompany antiphase domains, i.e., domains with phase shifts. When all antiphase domains are considered, those spin modulations always have unbroken **T** symmetry, just like commensurate even-period spin modulations.

**Magnetic point group analysis**

90 out of 122 MPGs have broken **PT**[39,40], and 31 out of these 90 MPGs belong to the ferromagnetic point group and do have broken **T** symmetry. When any of these 31 MPGs have fully compensated spins, then they belong to M-type altermagnets. 38 out of these 90 MPGs do not belong to the ferromagnetic point group but do have broken **T** symmetry. All these 38 MPGs belong to S-type altermagnets, since all of them do have fully compensated spins. There are 21 MPGs with unbroken **T**, broken **P**, and broken **PT** symmetry and all these 21 MPGs belong to A-type altermagnet, since all of them do have fully compensated spins. Among 38 S-type altermagnets, 27 can have broken **P**, **T** and **PT**, i.e. S/A type-altermagnets. These classifications are summarized in Fig. 2.

Figure 2 represents an expanded diagram compared to the one in our previous report[8]. The updated analysis differs from our earlier work due to a revised approach that now includes all directions, rather than focusing on the principal axes (high-symmetry axes) aligned with the basis vectors of conventional crystallographic coordinate systems[41]. For example, in our previous report, the MPG analysis was restricted to symmetries along the *x*, *y*, and *z* directions for orthorhombic MPGs. The current analysis, however, incorporates symmetries for allowed properties of altermagnets along all directions. This refined classification provides a more inclusive and systematic consideration of symmetries and physical properties along various directions. Both approaches are useful in different manners – in terms of simplicity, the approach in this paper may be better, but in terms of performing experiments, that in Ref. 8 can be useful, since experiments tend to be performed along symmetric directions, and the magnitudes of the relevant effects tend to be large along symmetric directions.

*P*-wave magnetism is a subset of A-type altermagnetism in our classification. It was proposed for CeNiAsO with non-collinear spins and the MPG of **21′**.[27,42,43] In A-type altermagnets, spin splitting in A-type altermagnets is antisymmetric with respect to *k* due to broken **P** symmetry. This splitting can occur without SOC and arises through 'magnetic order'. While this *k*- antisymmetric spin splitting in A-type altermagnets is sometime Rashba-type, it is not always the case (see Supplementary Note 2). Additional examples of ***mmm*** (S-type, MnTe)[15,44,45], ***6′mm′*** (S/A-type, $Fe_2Mo_3O_8$)[46], and ***6m′m′*** (M-type, $Mn_2Mo_3O_8$)[47] are discussed for their physical properties in the Supplementary Note 1. The MPGs ***6′mm′*** and ***6m′m′*** are particularly interesting due to broken all of **P**, **T**, and **PT**, leading to a variety of physical properties and phenomena.

**SPIN ROTATION OPERATION**

Broken **PT** symmetry alone doesn't allow us to determine whether the spin spitting of bands is due to SOC, which is a relativistic effect, or if it can still exist in the non-relativistic limit (i.e. for zero SOC). We now introduce a Spin Rotation Operation $S_n(r)$ around the $r$ axis, which rotates all spins at their own locations by $2\pi/n$ without rotating the crystallographic structure. When SOC is turned off, this $S_n(r)$ does not cost any energy, but this $S_n(r)$ symmetry may or may not exist in a given magnetic state along a given axis. Note that the presence of this $S_n(r)$ symmetry can vary even among magnetic states with the identical magnetic point group. This $S_n(r)$ in real space accompanies $\sigma(r)$ (spin expectation value along $r$ in crystal momentum space) without changing $k$.

In Fig. 1a, $S_2(x)$ and $S_2(z)$ link '+$k$ with spin up' and '+$k$ with spin down'. This means that the requirement for a system to have a spin-split bands for zero SOC is, at most, one unbroken $S_n(r)$ in addition to broken **PT** symmetry. Having more than one axis for $S_n(r)$ invariance prevents strong altermagnetism, making the absence of such symmetry a necessary condition for the emergence of strong altermagnetism. For zero SOC, a system with unbroken $S_2(y)$ symmetry, as shown in Fig. 1b, cannot have $\sigma(x)$ and $\sigma(z)$, but may have $\sigma(y)$. The spin-split band and symmetry are demonstrated in Supplementary Note 2. The procedures of considering spin rotation operations for exemplary magnetic states are illustrated in the Supplementary Note 3.

**STRONG v.s. WEAK ALTERMAGNETS**

With the concept of spin rotation operation, now, we can distinguish strong vs. weak altermagnetism: strong altermagnets do have spin-split bands through exchange coupling in the non-relativistic limit, i.e. for zero SOC, and weak altermagnets can have spin-split bands only with non-zero SOC. Note that the definition of strong and weak altermagnetism defined here differ from those in a recent report[48]. For orthogonal $x$, $y$ and $z$ axes, a magnetic system 'cannot' have spin-split bands for zero SOC if it has two or more spin rotation symmetries out of $S_n(x)$, $S_n(y)$, and $S_n(z)$, since it accompanies the presence of two or more spin rotation symmetries out of $\sigma(x)$, $\sigma(y)$, and $\sigma(z)$ in $k$ space[20]. The orthogonal $x$, $y$ and $z$ axes are along any directions, not just along crystallographic symmetric directions. For example, the ferromagnetic state in Fig. 1b has only $S_n(y)$, so it can have spin-split bands with $\sigma(y)$ for zero SOC, but the antiferromagnetic state in Fig. 1c has all of $S_n(x)$, $S_n(y)$, and $S_n(z)$ symmetries, so cannot have spin-split bands for zero SOC.

M-type altermagnet in Fig. 1d & 1e, and S-type altermagnet in Fig. 1g have $S_2(x)$, $S_2(y)$, and $S_2(z)$ symmetries, respectively, so they can have spin-split bands through exchange coupling with $\sigma(x)$, $\sigma(z)$, and $\sigma(y)$, respectively, for zero SOC. M-type altermagnet in Fig. 1f has no $S_n(x,y,z)$ symmetry, so can have spin-split bands with any of $\sigma(x)$, $\sigma(y)$, and $\sigma(z)$ for zero SOC. Thus, all of Figs. 1d-1g with broken **T** symmetry are strong altermagnets. It turns out that A-type altermagnet in Fig. 1h has $S_2(z)$ symmetry, so can have spin-split bands with $\sigma(z)$ for zero SOC and A-type altermagnet in Fig. 1i has $S_2(z)$ symmetry, so can have spin-split bands with $\sigma(z)$ for zero SOC. However, A-type altermagnet in Fig. 1j has all of $S_2(x)$, $S_2(y)$, and $S_2(z)$, so cannot have spin-split bands for zero SOC. Thus, A-type altermagnets in Fig. 1h & i are strong altermagnets and A-type altermagnet in Fig. 1j is a weak altermagnet. Since electric polarizations are associated with both the strong altermagnet in Fig. 1h and the weak altermagnet in Fig. 1j, their altermagnetic spin-split bands can be switched by external electrics fields.

The procedures of considering spin rotation operations for exemplary magnetic states are illustrated in the Supplementary Note 3. We emphasize that the spin structures in Fig. 1h and 1j have basically the same MPG, but that in Fig. 1h is a strong altermagnet, and that in Fig. 1j is a weak altermagnet, which clearly demonstrates that MPG itself as discussed in our previous work[8] is not sufficient to identify strong vs. weak altermagnets. All strong collinear altermagnets discussed in the original proposal[3,5] in the literature are a subset of our M-type and S-type strong altermagnets. M-type strong altermagnets can exhibit spin-split bands without SOC but acquire net orbital moments when SOC is present. A-type strong altermagnets can display spin-split bands without SOC while maintaining time-reversal invariance.

In the cases of M-type and S-type altermagnets with broken **T** symmetry and "collinear" spins, we can prove that they are always strong altermagnets. When the collinear spins are along $z$, $S_2(x)$ and $S_2(y)$ are identical with **T** symmetry operation, so they are broken, even though $S(z)$ is unbroken. Thus, they are strong altermagnets and can have spin-split bands with $\sigma(z)$ for zero SOC. Likely, the same argument works even for M-type and S-type altermagnets with broken **T** symmetry and non-collinear spins, but the exact proof seems depending on the magnetic unit cell.

Now, we can prove that all 'collinear' A-type altermagnets with unbroken **T** symmetry and broken **P** symmetry such as the polar magnetic state in Fig. 1j are always weak. When the collinear spins are along $z$, $S_2(x)$ and $S_2(y)$ are identical with **T** symmetry operation, so they are unbroken, in addition to unbroken $S(z)$. Thus, they are always weak. However, some 'non-collinear' A-type altermagnets with unbroken **T** symmetry and broken **P** symmetry such as the polar magnetic state in Fig. 1h and the chiral magnetic state in Fig. 1i are strong, as discussed earlier.

**STRONG NONCOLLINEAR ALTERMAGNETS ON KAGOME LATTICE**

Kagome lattice turns out to be a wonderful system where all three types of altermagnetism can be realized. All of Figs. 3 have fully compensated spins on Kagome lattice. Fig. 3c has unbroken **PT** while the rest of Fig. 3 do have broken **PT**. Figs. 3a-3b have broken **T**, and do have net magnetic moments (along $y$, and $x$, respectively). Figs. 3d-3f have broken **T**, and do not have net magnetic moments. Fig. 3f has broken all of **P**, **T** and **PT**, and Fig. 3g has unbroken **T** and broken **P**. Thus, Fig. 3a-b, Fig. 3d-e, Fig. 3f, and Fig. 3g are examples of M-type, S-type, S/A-type and A-type altermagnets on Kagome lattice, respectively. Fig. 3a with ***m'mm'*** can have a net magnetic moment along $y$ and is realized in $Mn_3Ge(Ga)$[49,50], and Fig. 3b with ***mm'm'*** can have a net magnetic moment along $x$ and is realized in $Mn_3Sn$[16,29,51]. All of these M-type altermagnets can exhibit linear anomalous Hall effects in the planes perpendicular to the net magnetic moment directions. S-type altermagnet on Kagome lattice shown in Fig. 3d with ***6'/m'm'm***, which can happen in each (111) layer of cubic $Mn_3Ir(Pt,Rh)$[52], exhibit transverse even-order current-induced magnetization with current along $x$ and induced magnetization along $y$, high-odd-order AHE with current along $x$ and Hall voltage along $z$, and transverse piezomagnetism with uniaxial stress along $x$ and induced magnetization along $y$.

Fig. 3f with $\bar{6}'m'2$ can exhibit transverse even-order current-induced magnetization with current along $x$ and induced magnetization along $y$, high-odd-order AHE with current along $x$ and Hall voltage along $z$, and transverse piezomagnetism with uniaxial stress along $x$ and induced magnetization along $y$. A-type altermagnet in Fig. 3g with ***m1'*** can have electric polarization in the $xy$ plane. This A-type altermagnet can exhibit even-order AHE with applied current perpendicular to the electric polarization and Hall voltage along the electric polarization. All of these altermagnets on Kagome lattice in Fig. 3a-3b, 3d-3g have, at most, $S_2(z)$ symmetry, so they are strong altermagnets and can have spin-split bands with, at least, $\sigma(z)$ for zero SOC.

**CONCLUSION**

The definition of Altermagnets is refined: magnetism with broken **PT** symmetry and complete spin compensation in the non-relativistic limit. – here, collinearity or non-collinearity is not important. Three kinds of altermagnets are identified: M-type: broken **T** symmetry and nonzero net magnetic moments, S-type: broken **T** symmetry, and zero net magnetic moments, and A-type: unbroken **T** and broken **P** symmetries. S-type altermagnets do not belong to the ferromagnetic point group, and **P** may or may not be broken in S-type altermagnets. M-type altermagnets, belong to the ferromagnetic point group, show linear AHE, transverse even-order current induced magnetization, and transverse piezomagnetism. S-type altermagnets do exhibit high odd-order AHE, transverse even-order current induced magnetization, and transverse piezomagnetism. A-type altermagnets allows quadratic AHE, transverse odd-order current induced magnetization, and transverse piezoelectricity. In fact, all altermagnets exhibit some form of kinetomagnetism, i.e. current-induced magnetization in various orders.

We also classified strong vs. weak altermagnets: strong altermagnets have spin-split bands through exchange coupling even for zero SOC, and weak altermagnets have spin-split bands only for non-zero SOC. The total number of symmetric orthogonal spin rotation operations determines if an altermagnet is strong or weak. All 'collinear' M-type and S-type altermagnets are strong altermagnets, and all 'collinear' A-type altermagnets are weak altermagnets. These clear and comprehensive classifications of altermagnets based on the **PT** symmetry approach provide a far-reaching perspective for future research on altermagnetism.

**ACKNOWLEDGEMENTS:** This work was supported by the DOE under Grant No. DOE: DE-FG02-07ER46382.
**AUTHOR CONTRIBUTIONS:** S.W.C. conceived and supervised the project. F.-T.H. conducted magnetic point group analysis. S.W.C. wrote the remaining part.
**DATA AVAILABILITY:** All study data is included in the article.

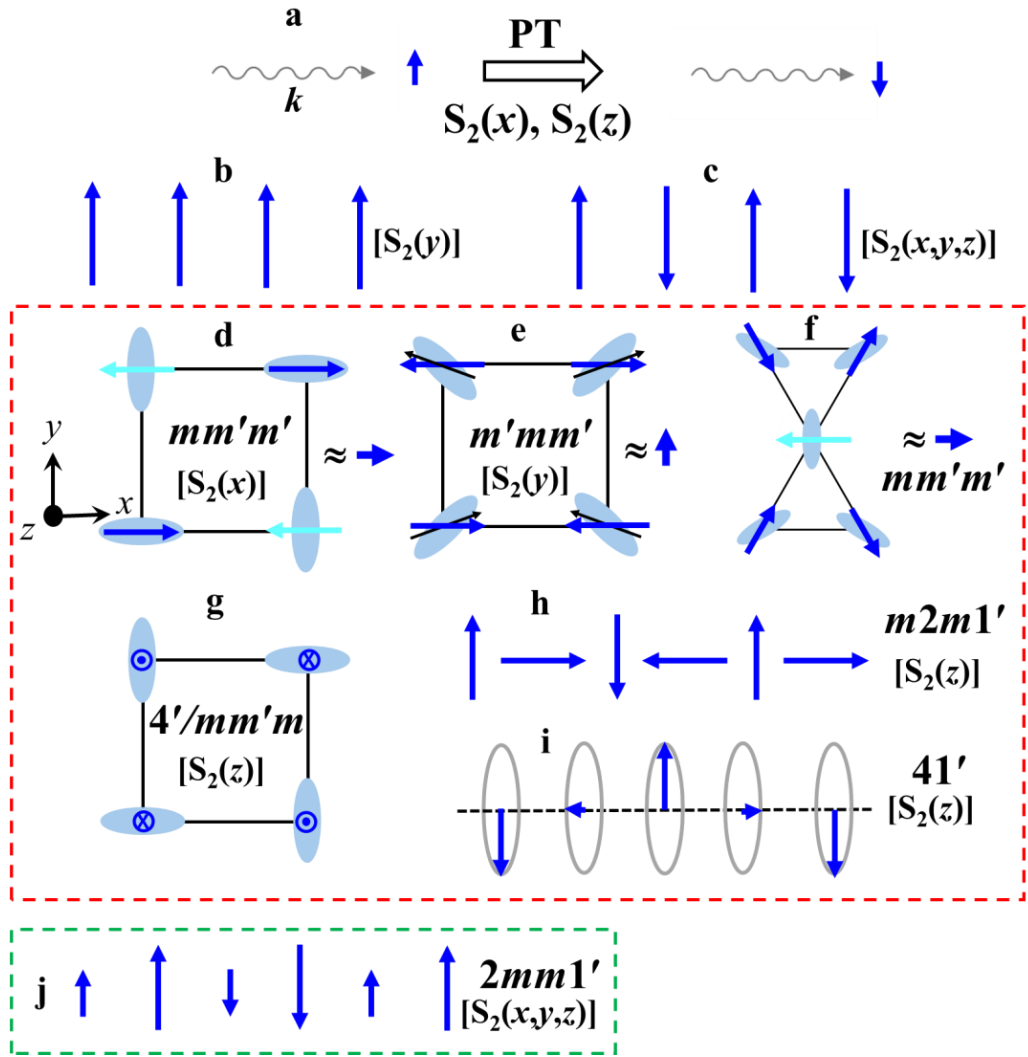

**Fig. 1 Various spin configurations. a**, PT, $S_2(x)$, or $S_2(z)$ symmetry operations link between 'electron crystal momentum $k$ with up spin' and '$k$ with down spin'. **b**, ferromagnetic spins. **c**, antiferromagnetic spins. **b&c** are not altermagnets. **d**, in-plane collinear spins with alternating local structures on square lattice. **e**, in-plane collinear spins with alternating local structures on square lattice. **f**, an alternating arrangement of three distinct crystallographic-director orientations and three spin orientations, which can be realized in Kagome lattice (see Fig. 3b). **g**, out-of-plane collinear spins with alternating local structures on square lattice. Blue arrows are spins, black arrows indicate canted magnetic moments due to the DM interaction, and dark/light blue colors of blue arrows depict different orbital moments due to different g-tensor anisotropies. **h**, cycloidal spins. **i**, helical spins. **j**, up-up-down-down collinear two kinds of spins. $[S_n(x,y,z)]$ indicates the presence of all of $S_n(x)$, $S_n(y)$, and $S_n(z)$, meaning no spin-split bands occur for zero SOC, i.e. weak altermagnetism (green-dashed-box). [$S_n(x)$] means that only the $S_n(x)$ symmetry is present, allowing for spin-split bands with σ(x) even for zero SOC, i.e. strong altermagnetism (red-dashed-box). M-type altermagnet: **d-f**; S-type altermagnet: **g**; and A-type altermagnets: **h-j**

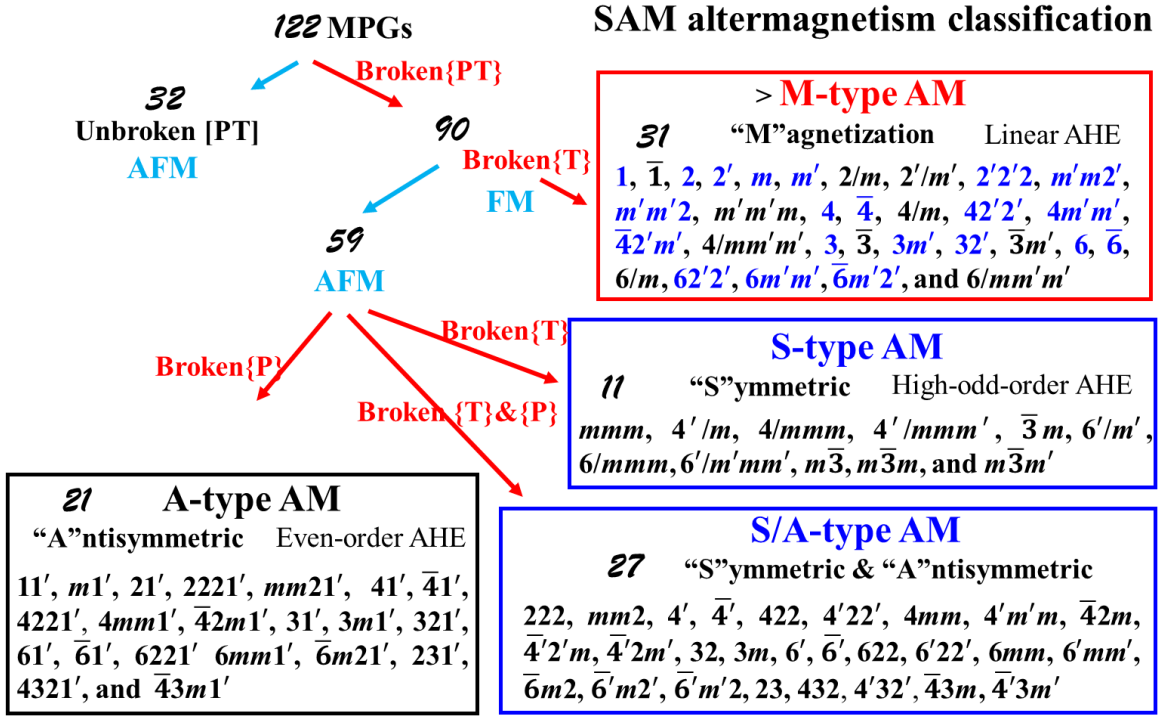

**Fig. 2 MPG classification for M-type, S-type, and A-type altermagnets.** 31 MPGs belonging to the ferromagnetic point group can be M-type AM if they have fully compensated spins, there are 11 MPGs for S-type AM, 27 MPGs for S/A-type, and there exist 21 MPGs for A-type AM. Square brackets [ ] denote unbroken symmetry, while curly brackets { } represent broken symmetry. All altermagnets have broken **PT**, M-type and S-type AM have broken **T**, S/A-type has broken **T** and **P**. A-type altermagnets have unbroken **T** and broken **P**. MPGs in blue and S/A-type AM have broken all of **P**, **T**, and **PT**. M-type altermagnets show non-zero net magnetization (i.e., linear AHE), S-type altermagnets can have transverse even-order current-induced magnetization (i.e., high-odd-order AHE), and A-type altermagnets can exhibit transverse odd-order current-induced magnetization (i.e., even-order AHE). Thus, both M-type and S-type altermagnets can have transverse piezomagnetism. MPGs in blue for M-type and S/A-type altermagnets can also show transverse odd-order current-induced magnetization (i.e., even-order AHE).

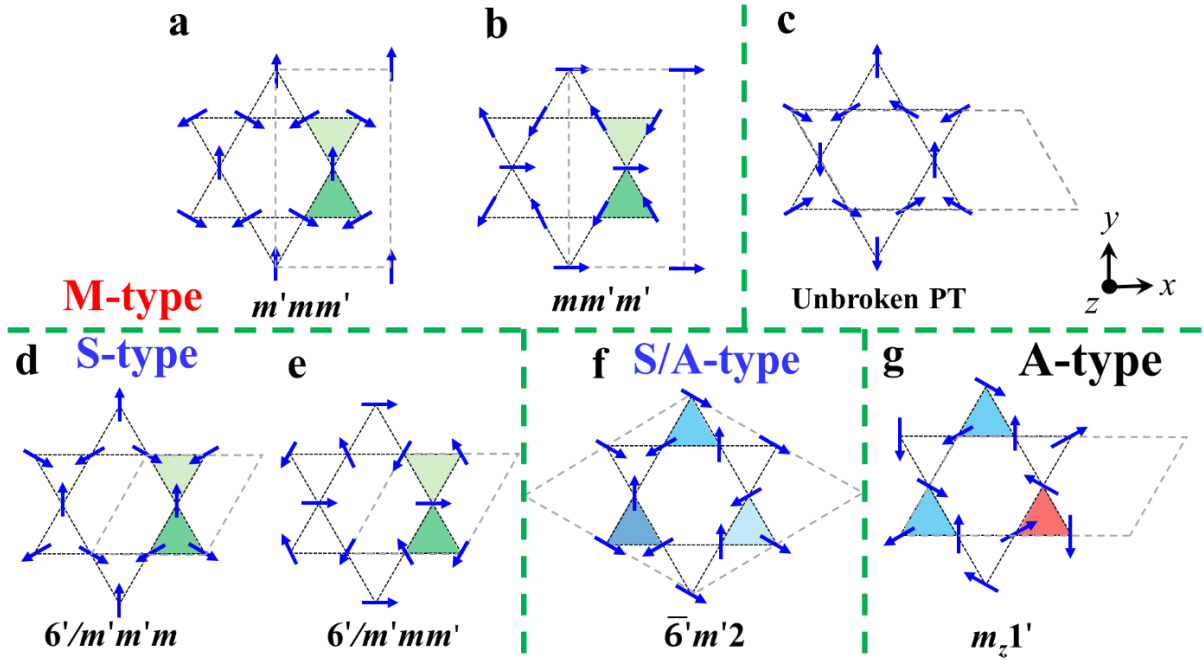

**Fig. 3 Various spin configuration on Kagome lattices. a-b**, M-type altermagnets. **c**, a spin configuration with unbroken **PT**. **d-f**, S-type altermagnets. **g**, A-type altermagnet. All configurations except **c** have broken **PT**. **a**, **b**, and **d-e** have broken **T** and unbroken **P**. Green and light green units indicate inversion relation. **f** has broken all of **P**, **T** and **PT** with blue units representing a three-fold relationship. **g** has unbroken **T** and broken **P**, with blue and red units representing time-reversal relation. Grey dashed lines outline the magnetic unit cells.

# Altermagnetism Classification

Sang-Wook Cheong[*1] and Fei-Ting Huang[1]

[1]Keck Center for Quantum Magnetism and Department of Physics and Astronomy, Rutgers University, Piscataway, NJ 08854, USA

**Supplementary Note 1: Physical properties in M-type and S-type altermagnets**

Herein, we discuss physical properties and phenomena of ***mmm*** (S-type altermagnet, MnTe), **6′*mm*′** (S/A-type altermagnet, $Fe_2Mo_3O_8$), and **6*m*′*m*′** (M-type altermagnet, $Mn_2Mo_3O_8$). Note that both **6′*mm*′** and **6*m*′*m*′** have broken all of **P**, **T**, and **PT**, so exhibit a variety of physical properties and phenomena.

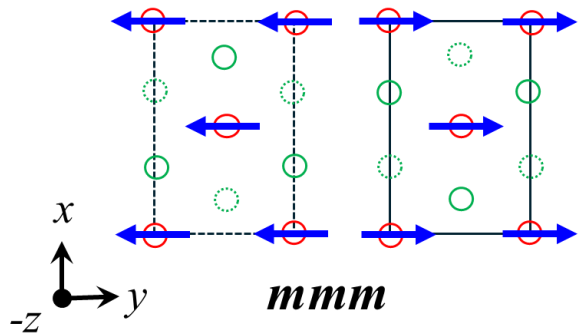

**Supplementary Figure 1** The magnetic state of ***mmm***, which can be realized in MnTe. ***mmm*** is a S-type altermagnet, and can exhibit transverse even-order current-induced magnetization along *z* with current along *xy* or *yx*, high-odd-order AHE with current along *xy* (*yx*), and Hall voltage along *yx* (*xy*), and transverse piezomagnetism with stress along *xy* or *yx* and induced magnetization along *z*.

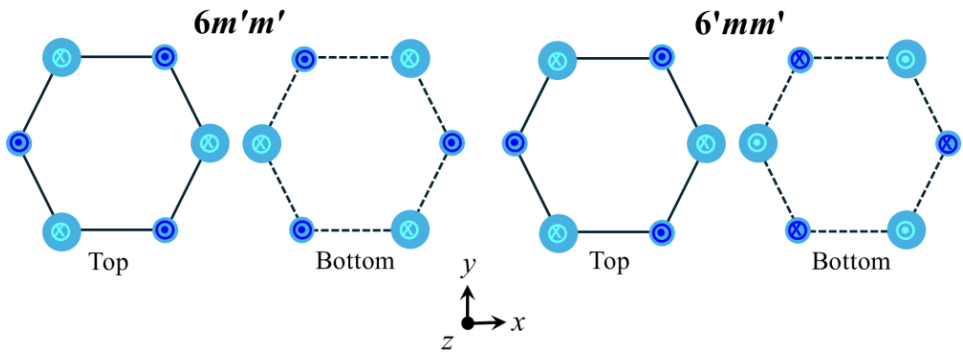

**Supplementary Figure 2** MPG **6*m*′*m*′** is a M-type altermagnet and MPG **6′*mm*′** is a S/A-type altermagnet, and both **6*m*′*m*′** and **6′*mm*′** have broken all of **P**, **T** and **PT**. The magnetic state of **6*m*′*m*′** can be realized in $Mn_2Mo_3O_8$ and the magnetic state of **6′*mm*′** can be realized in $Fe_2Mo_3O_8$.

**6m'm'** exhibit a net electric polarization along *z*, a net magnetic moment along *z*, linear AHE with current along *x* (*y*) and Hall voltage along *y* (*x*), transverse odd-order current-induced magnetization with current along *x* (*y*) and induced magnetization along *y* (*x*), and even-order AHE with current along *x* or *y* and Hall voltage along *z*. **6'mm'** has a net electric polarization along *z*, and can exhibit transverse odd-order current-induced magnetization with current along *x* (*y*) and induced magnetization along *y* (*x*), even-order AHE with current along *x* or *y* and Hall voltage along *z*, transverse piezomagnetism with stress along *y* and induced magnetization along *x*, transverse even-order current-induced magnetization with current along *y* and induced magnetization along *x*, and high-odd-order AHE with current along *y* and Hall voltage along *z* (the first two effects are due to the presence of electric polarization, and the last two effects are related with transverse piezomagnetism). $Mn_2Mo_3O_8$ and $Fe_2Mo_3O_8$ are insulating, so the relevant effect is anomalous thermal Hall effect, rather than AHE.

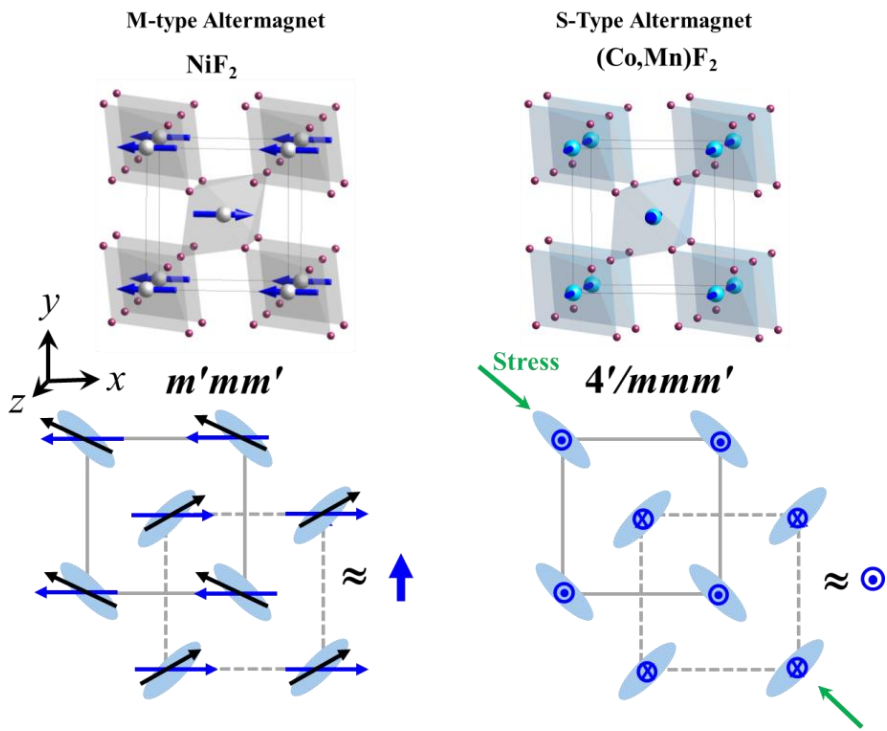

**Supplementary Figure 3** $NiF_2$ with MPG **m'mm'** is a M-type altermagnet and $(Co,Mn)F_2$ with MPG **4'/mmm'** is a S-type altermagnet. $NiF_2$ is a classic example of weak ferromagnets induced by SOC, and exemplifies a M-type (strong) altermagnet as shown below. The blue arrows represent the in-plane spin (left panel) structure in $NiF_2$ with crystallographic space group **P4₂/mnm** and magnetic point group (MPG) **m'mm'**. In the spin configuration, spins are fully compensated. However, Dzyaloshinskii-

Moriya (DM) interaction induces canting of orbital moments (black arrows), thus leading to a net orbital moment and weak ferromagnetism. Note that when a net orbital moment develops through canting, spins can be also canted through SOC (a secondary effect), so spins can contribute to the total net magnetic moment. When the spins are aligned along $z$ directions with the same crystallographic structure as $NiF_2$, the system turns to a S-type strong altermagnet (MPG, **4′/mmm′**) as seen in $CoF_2$ or $MnF_2$ (right panel), which are known to be transverse piezomagnetic materials.

## Supplementary Note 2: Spin-split bands and symmetry

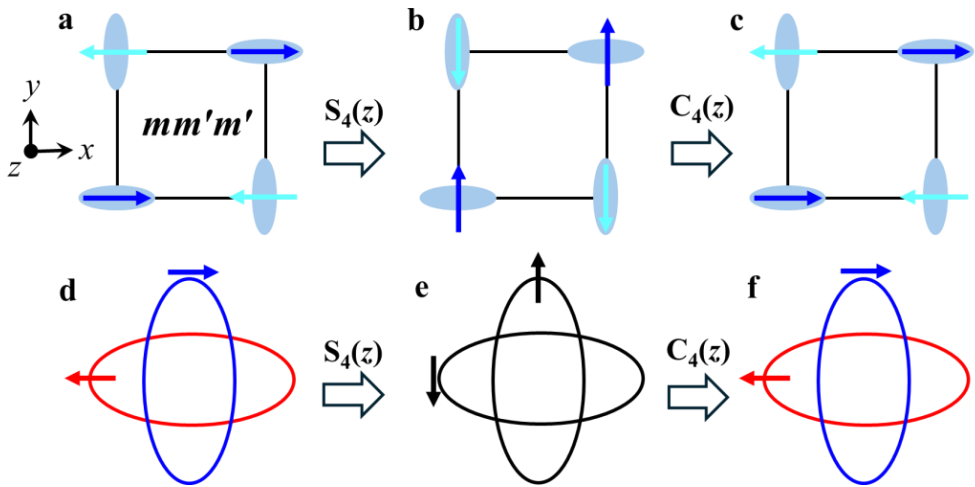

**Supplementary Figure 4 Symmetry consistency between the collinear spin configuration of a strong M-type altermagnet and its spin-split bands.** The spin configuration in Fig. 1d has unbroken [$S_2(x)$] and broken {$S_n(y),S_n(z)$}, so non-zero $\sigma_x$ can exist in the non-relativistic limit (i.e., a strong altermagnet). This spin configuration has $C_4(z) \otimes S_4(z)$ symmetry as well as other symmetries such as [$C_{2x}$]. The spin-split band in **d** is consistent with these symmetries.

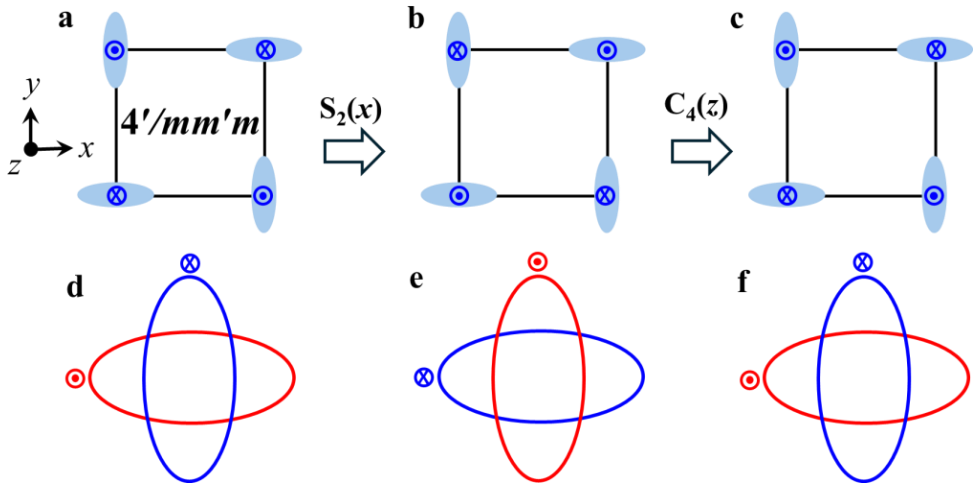

**Supplementary Figure 5 Symmetry consistency between the collinear spin configuration of a strong S-type altermagnet and its spin-split bands.** The spin configuration in Fig. 1g has unbroken [$S_2(z)$] and broken {$S_n(x), S_n(y)$}, so non-zero $\sigma_z$ can exist in the non-relativistic limit (i.e., a strong altermagnet). This spin configuration has $C_4(z) \otimes S_2(x)$ symmetry as well as other symmetries such as [$C_{2z}$]. The spin-split band in **d** is consistent with these symmetries.

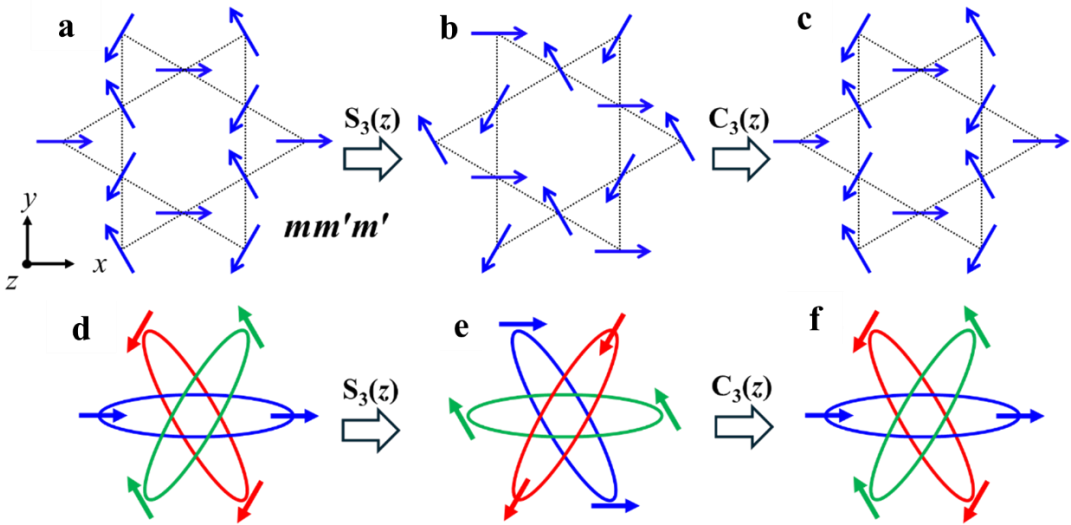

**Supplementary Figure 6 Symmetry consistency between the non-collinear spin configuration of a strong M-type altermagnet and its spin-split bands.** The spin configuration similar in Fig. 3a has broken {$S_n(x), S_n(y), S_n(z)$}, so non-zero any of $\sigma_x$, $\sigma_y$, and $\sigma_z$ can exist in the non-relativistic limit (i.e., a strong altermagnet). This spin configuration has $C_3(z) \otimes S_3(z)$ symmetry as well as other symmetries such as [$C_{2x}$]. The spin-split band in **d** is consistent with these symmetries.

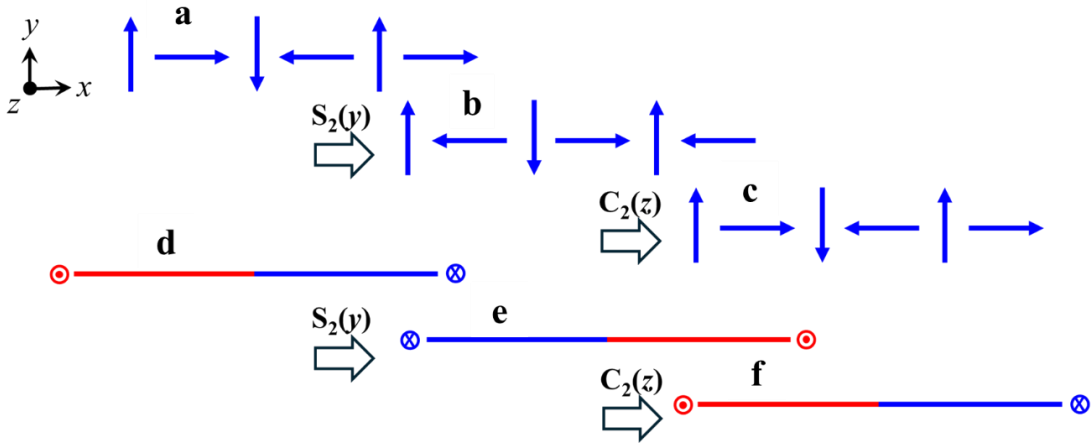

**Supplementary Figure 7 Symmetry consistency between the non-collinear spin configuration of a strong A-type altermagnet and its spin-split bands.** The spin configuration in Fig. 1h has unbroken [$S_2(z)$}, so non-zero $\sigma_z$ can exist in the non-relativistic limit (i.e., a strong altermagnet). This spin configuration has $C_2(z) \otimes S_2(y)$ symmetry as well as other symmetries such as [$C_{2y}$]. The spin-split band in **d** is consistent with these symmetries, and antisymmetric with respect to $k$. This splitting is similar with the Rashba splitting, but exists even for zero SOC.

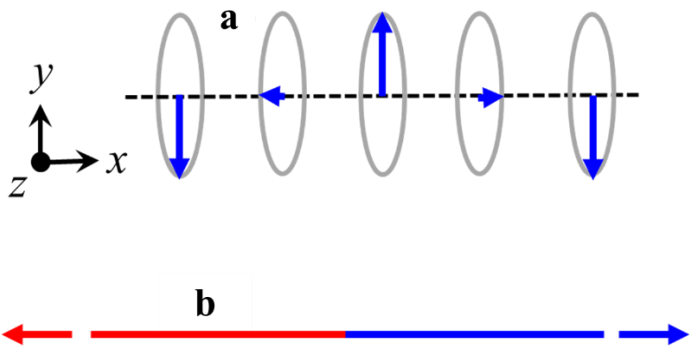

**Supplementary Figure 8 Symmetry consistency between the non-collinear spin configuration of a strong A-type altermagnet and its spin-split bands.** The spin configuration in Fig. 1i has unbroken [$S_2(z)$], so non-zero $\sigma_z$ can exist in the non-relativistic limit (i.e., a strong altermagnet). This spin configuration has unbroken [$C_{2x}, C_{2y}, C_{2z}$], so it does not have the non-trivial invariance of the type of spin rotation operation times spatial rotation operation. The spin-split band in **b** is consistent with unbroken [$C_{2x}, C_{2y}, C_{2z}$] symmetries, and antisymmetric with respect to $k$ (but, different from the Rashba-type), and exists without SOC.

**Supplementary Note 3: Spin rotation operation**

Spin Rotation Operation of $S_n(x)$ is defined for rotating all spins at their own locations by $2\pi/n$ around x-axis without rotating the crystallographic structure. When spin-orbit coupling (SOC) is turned off, this $S_n(x)$ does not cost any energy, but a given magnetic system may or may not have this $S_n(x)$ symmetry. When this $S_n(x)$ symmetry is present in a magnetic state, then electronic bands in $k$ space for zero SOC can have neither $\sigma(y)$ nor $\sigma(z)$ but may have $\sigma(x)$. For orthogonal $x$, $y$ and $z$ axes, a magnetic system "cannot" have spin-split bands for zero SOC if it has two or more spin rotation symmetries out of $S_n(x)$, $S_n(y)$, and $S_n(z)$, since it accompanies the presence of two or more spin rotation symmetries out of $\sigma(x)$, $\sigma(y)$, and $\sigma(z)$ in $k$ space. If so, then it can be only a weak altermagnet. Note that orthogonal $x$, $y$ and $z$ axes are along any directions, not just along crystallographic symmetric directions. A few examples of how to identify strong vs. weak altermagnets with symmetry consideration of Spin Rotation Operation are illustrated below:

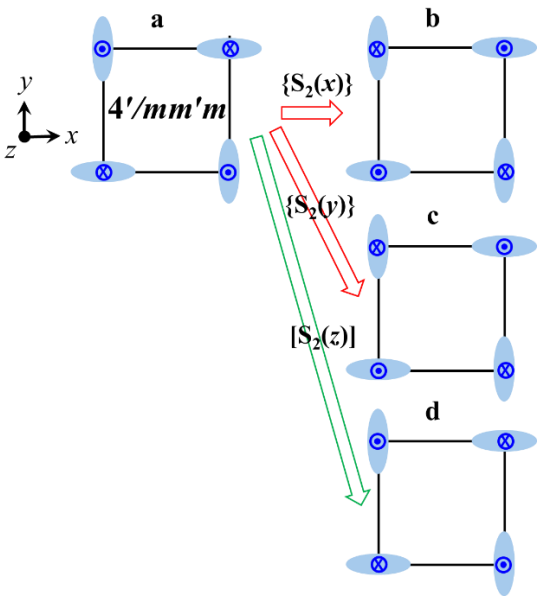

**Supplementary Figure 9 Strong S-type altermagnet with collinear spins. a,** The magnetic state exhibiting symmetry **4′/mm′m**. **b-d**, The spin states through spin rotations at each site along $x$, $y$, and $z$ axes, respectively. These configurations reveal broken $S_2(x)$ and $S_2(y)$, and unbroken $S_2(z)$. This indicates that the magnetic state shown in **a** is a strong S-type altermagnet.

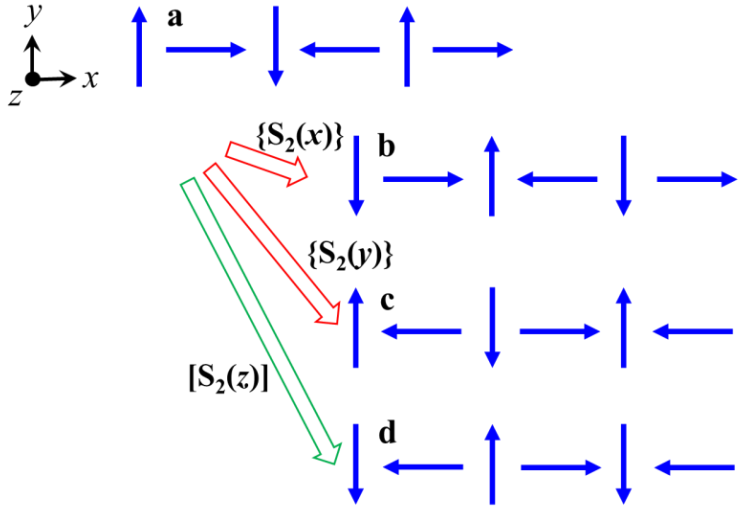

**Supplementary Figure 10 Strong A-type altermagnet with noncollinear spins. a,** The magnetic state exhibiting symmetry ***m2m1'***. **b-d**, The spin states through spin rotations at each site along $x$, $y$, and $z$ axes, respectively. These configurations reveal broken $S_2(x)$ and $S_2(y)$, and unbroken $S_2(z)$ (and also unbroken $S_4(z)$). This indicates that the magnetic state shown in **a** is a strong A-type altermagnet.

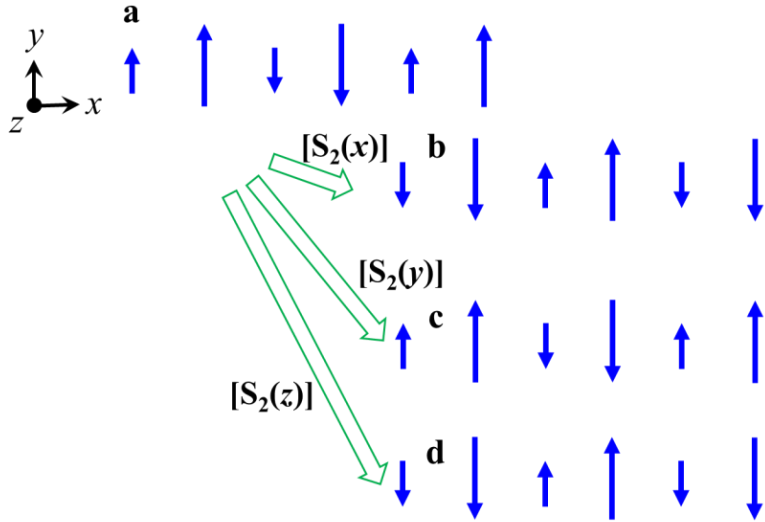

**Supplementary Figure 11 Weak A-type altermagnet with collinear spins.** The magnetic state exhibiting symmetry ***2mm1'***. **b-d**, The spin states through spin rotations at each site along $x$, $y$, and $z$ axes, respectively. These configurations reveal unbroken $S_2(x)$ and $S_2(y)$, and $S_2(z)$. This indicates that the magnetic state shown in **a** is a weak A-type altermagnet.

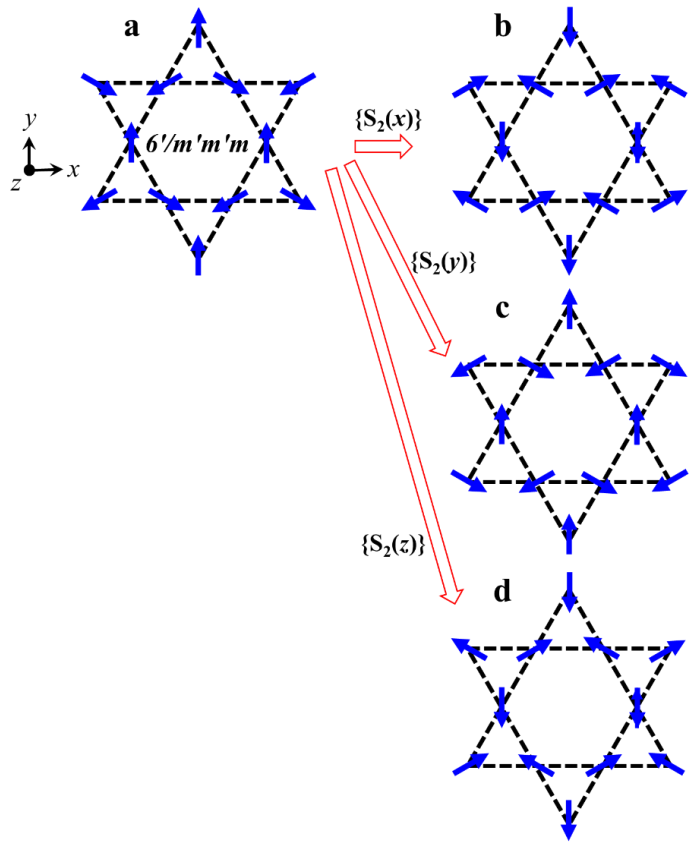

**Supplementary Figure 12 Strong S-type altermagnet with noncollinear spins.** The magnetic state exhibiting symmetry **6′/m′m′m**. **b-d**, The spin states through spin rotations at each site along $x$, $y$, and $z$ axes, respectively. These configurations reveal broken $S_2(x)$ and $S_2(y)$, and $S_2(z)$. This indicates that the magnetic state shown in **a** is a strong S-type altermagnet with noncollinear spins.